\documentclass{elsart}
\usepackage[dvips]{graphicx}
\begin{document}
\begin{frontmatter}

\title{Search for $\beta^+$EC and ECEC processes in $^{74}$Se}
\author[ITEP]{A.S.~Barabash$^{\;1)}$}\thanks{Corresponding author, 
Institute of Theoretical and Experimental Physics, B.~Cheremushkinskaya 25,
117259 Moscow, Russia,  e-mail: Alexander.Barabash@itep.ru,
tel.: 007 (495) 129-94-68, fax: 007 (495) 127-08-33},
\author[CENBG]{Ph.~Hubert},
\author[CENBG]{A.~Nachab},
\author[ITEP]{V.~Umatov}
\address[ITEP]{Institute of Theoretical and Experimental Physics, B.\
Cheremushkinskaya 25, 117259 Moscow, Russian Federation}
\address[CENBG]{Centre d'Etudes Nucl\'eaires,
IN2P3-CNRS et Universit\'e de Bordeaux, 33170 Gradignan, France}
\date{ }

\begin{abstract}
 For the first time, limits on double-beta processes in $^{74}$Se have been obtained using 
a 400 cm$^3$ HPGe detector and an external source consisting of natural selenium powder.
 At a confidence level of 90\%, they are $1.9\times 10^{18}$ y for the $\beta^+$EC$(0\nu + 2\nu)$ 
transition to the ground state, $7.7\times 10^{18}$ y for the ECEC($2\nu$) transition to the 
$2^+_1$ excited state in $^{74}$Ge (595.8 keV), $1.1\times 10^{19}$ y for the ECEC($0\nu$) transition to the 
$2^+_1$ excited state in $^{74}$Ge (595.8 keV) and $5.5\times 10^{18}$ y for the ECEC($2\nu$) 
and ECEC($0\nu$) transitions to the 
$2^+_2$ excited state in $^{74}$Ge (1204.2 keV). The last transition is discussed in association 
with a possible enhancement of the decay rate, in this case by several orders of magnitude, 
because the ECEC$(0\nu)$ process is nearly degenerate with an excited state in the daughter 
nuclide. Prospects for investigating such processes in future experiments are discussed.


\end{abstract}
\end{frontmatter}

\newpage

\section{Introduction}
The experiments with solar, atmospheric, reactor and accelerator neutrinos have provided compelling 
evidences for the existence of neutrino oscillations driven by nonzero neutrino masses 
and neutrino mixing (see recent reviews \cite{VAL06,BIL06,MOH06} and reference therein).
These results are impressive proof that neutrinos have a nonzero mass. However, the experiments 
studying neutrino oscillations are not sensitive to the nature of the neutrino mass (Dirac or Majorana) 
and provide no information on the absolute scale of the neutrino masses, since such experiments are 
sensitive only to the difference of the masses, $\Delta m^2$. The detection and study of $0\nu\beta\beta$ 
decay may clarify the following problems of neutrino physics (see discussions in \cite{PAS03,MOH05,PAS06}:
 (i) neutrino nature: 
whether the neutrino is a Dirac or a Majorana particle, (ii) absolute neutrino mass scale 
(a measurement or a limit on $m_1$), (iii) the type of neutrino mass hierarchy (normal, inverted, or 
quasidegenerate), (iv) CP violation in the lepton sector (measurement of the Majorana CP-violating phases).
     Let us consider main modes of $\beta^-\beta^-$ decay:

\begin{equation}
(A,Z) \rightarrow (A,Z+2) + 2e^{-} + 2\tilde \nu
\end{equation}

\begin{equation}
(A,Z) \rightarrow (A,Z+2) + 2e^{-} 
\end{equation}

Process (1) is a second-order process, which is not forbidden by any conservation law. 
The detection of this process furnishes information about nuclear matrix elements (NME) for $2\beta$ 
transitions, which makes it possible to test the existing models for calculating these NMEs and 
contributes to obtaining deeper insight into the nuclear physics aspect of the double-beta decay 
problem. 
It is expected that the accumulation of experimental information about $2\nu\beta\beta$ processes will 
improve the quality of the calculations of NMEs, both for $2\nu$ and $0\nu$ decay. At the present 
time, $2\nu\beta\beta$ decay has so far been recorded for ten nuclei. In addition, the 
$2\nu\beta\beta$ decays of $^{100}$Mo and $^{150}$Nd to the first $0^+$ excited state of the 
daughter nuclides have been observed (see reviewes \cite{BAR06,BAR06a}).

Process (2) violates the law of lepton-number conservation ($\Delta$L =2) and requires that the 
Majorana neutrino has a nonzero rest mass or that an admixture of right-handed currents be present 
in weak interaction.  In contrast to two-neutrino decay, neutrinoless double-beta decay has not 
yet been observed\footnote{The possible exception is the result with $^{76}$Ge, published by a 
fraction of the  Heidelberg-Moscow Collaboration, $T_{1/2}\approx 1.2\cdot 10^{25}$ y \cite{KLP04}. 
The Moscow portion of the Collaboration does not agree with this conclusion 
\cite{BAK05} and this result was subjected to criticism in other papers (see, for example,  
\cite{STR05}).  Thus at the present time this ``positive'' result is not accepted by the 
``$2\beta$ decay community'' and it has to be checked by new experiments.} although, from the 
experimental point of view, it is easier to detect it. In this case, one seeks a peak in the 
experimental energy spectrum in the range of the double-beta transition energy with width 
determined by the detector's resolution. Only limits on the level of $\sim 10^{24} - 10^{25}$ y 
for half-lives and $\sim 0.35-1.3$ eV for effective Majorana neutrino mass $\left<m_\nu\right>$ 
have been obtained in the best modern experiments (see reviewes \cite{BAR06,BAR06a}).

 Much less attention has been given to the investigation of $2\beta^+$, $\beta^+$EC and ECEC 
processes although such attempts were done from time to time in the past (see review 
\cite{BAR04}). Again, the main interest here is connected with neutrinoless decay:

\begin{equation}
(A,Z) \rightarrow (A,Z-2) + 2e^{+} 
\end{equation}

\begin{equation}
e^- + (A,Z) \rightarrow (A,Z-2) + e^{+} + X
\end{equation}

\begin{equation}
e^- + e^- + (A,Z) \rightarrow (A,Z-2)^* \rightarrow (A,Z-2)+\gamma + 2X
\end{equation}

Process (3) has a very nice signature because, in addition to two positrons, four annihilation 511 
keV gamma quanta will be detected. On the other hand, the rate for this process should be much 
lower in comparison with $0\nu\beta\beta$ decay because of substantially lower kinetic energy 
available in such a transition (2.044 MeV is spent for creation of two positrons) and of the 
Coulomb barrier for positrons.  There are only 6 candidates for this type of decay: $^{78}$Kr, 
$^{96}$Ru, $^{106}$Cd, $^{124}$Xe, $^{130}$Ba and $^{136}$Ce. The half-lives of most prospective 
isotopes are estimated to be $\sim 10^{27} -10^{28}$ y (for $\left<m_\nu\right> = 1$ eV) 
\cite{HIR94}; this is approximately $10^3-10^4$ times higher than for $0\nu\beta\beta$ decay for 
such nuclei as $^{76}$Ge, $^{100}$Mo, $^{82}$Se and $^{130}$Te (see review \cite{BAR04}).

Process (4) has a nice signature (positron and two annihilation 511 keV gammas) and is not as 
strongly suppressed as $2\beta^+$ decay. In this case, half-life estimates for the best nuclei 
give $\sim 10^{26} - 10^{27}$ y (again for $\left<m_\nu\right> = 1$ eV) \cite{HIR94}.

In the last case (process (5)), the atom de-excites emitting two X-rays and the nucleus de-excites 
emitting one $\gamma$-ray (bremsstrahlung photon)\footnote{In fact processes with irradiation 
of inner conversion electron, $e^+e^-$ pair or two gammas are also possible \cite{DOI93} (in 
addition, see discussion in \cite{SUJ04}). These possibilities are especially important in 
the case of ECEC$(0\nu)$ transition with capture of two electrons from K shell - in this case 
the transition with irradiation of one $\gamma$ is forbidden \cite{DOI93}}.
 For a transition to an excited state of the daughter nucleus, besides a bremsstrahlung photon, 
$\gamma$-rays are emitted from the decay of the excited state.  Thus, in this case, we again have 
a very nice observational signature for this process. The rate is practically independent of decay 
energy and increases with both decreasing bremsstrahlung photon energy and increasing Z 
\cite{SUJ04,VER83}.  The rate is quite low even for heavy nuclei, with $T_{1/2} \sim 10^{28} 
-10^{31}$  y \cite{SUJ04}. 
Nevertheless, as was mentioned many years ago, a resonant process is possible if the mass 
of the daughter atom (in which electrons from the K (or L) shell are transformed into an excited state) 
coincides (within $< 1$ keV) with the mass of the initial atom in its ground state 
\cite{WIN55,VOL82,BER83}\footnote{In Ref. \cite{SUJ04}, the resonance condition for transitions 
with a bremsstrahlung photon is $E_{brems}=Q_{res}=\mid E(1S,Z-2)-E(2P,Z-2)\mid $, i.e. when the 
photon energy becomes comparable to the $2P-1S$ atomic level difference in the final atom.}. 
Possible rate enhancements on the order of $\sim 10^5-10^6$ may be obtained \cite{SUJ04,BER83}. 
In Ref. \cite{BER83}, the resonant transition $^{112}$Sn$\rightarrow ^{112}$Cd (0$^+$; 1871 keV) was 
investigated theoretically and the estimated half-life was $T_{1/2}\sim 3\cdot 10^{24}$ y (for 
$\left<m_\nu\right> = 1$ eV). Hence, the sensitivity of this process to the neutrino mass is 
comparable to the sensitivity of neutrinoless double-beta decay (see above). In Ref. \cite{FRE05}, 
it is proposed to search for the $^{74}$Se$\rightarrow ^{74}$Ge (1204.2 keV) resonant transition, 
the decay scheme of which is shown in Fig. 1.  Taking into account that the atomic mass difference 
$\Delta M$ between $^{74}$Se and $^{74}$Ge is known to an accuracy of 2.3 keV (at one standard 
deviation), one can discuss possible resonant capture of two electrons from L shell for the 
transition to the 1204.2-keV level in $^{74}$Ge. In 66\% of cases this process will be accompanied 
by a cascade of two $\gamma$-quanta of energies 608.4 and 595.8 keV; in 34\% of cases it will be 
one $\gamma$-quantum of energy 1204.2 keV.  It has to be mentioned that observation of these 
$\gamma$-quanta will provide extremely strong evidence of neutrinoless double-electron capture 
because two neutrino capture is heavily suppressed (because of the extremely low transition 
energy).

\begin{figure*}
\begin{center}
\resizebox{0.75\textwidth}{!}{\includegraphics{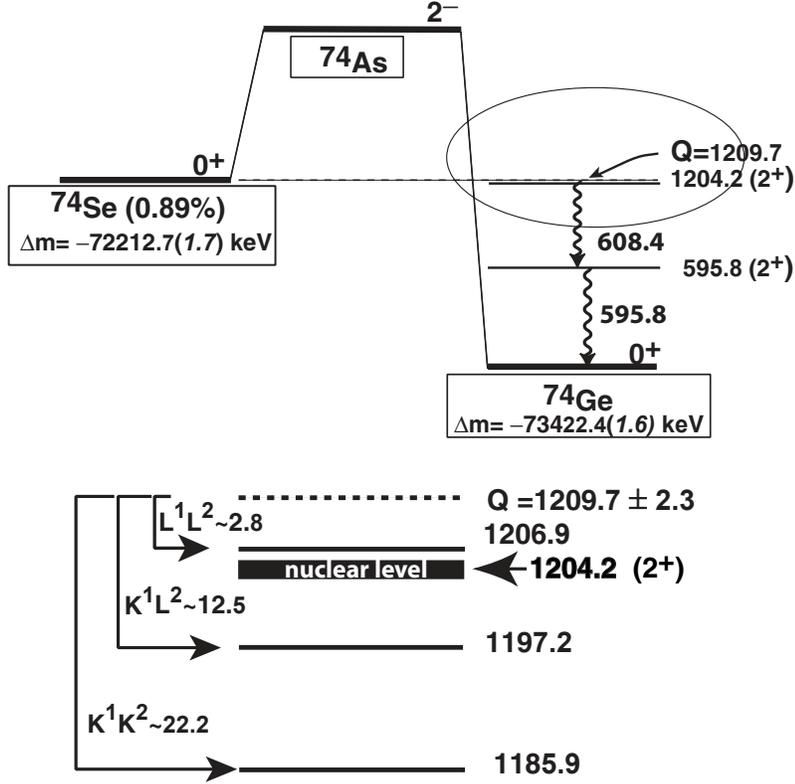}}
\caption{Energetics of the $^{74}$Se $0\nu$ECEC decay indicating the near degeneracy of the $^{74}$Se
ground state and the second excited state in $^{74}$Ge. The circle marks the part magnified in the 
lower part of the figure.}  
\label{fig_1}
\end{center}
\end{figure*}

For completeness, let us present the two-neutrino modes of $2\beta^+$, $\beta^+$EC and ECEC 
processes:

\begin{equation}
(A,Z) \rightarrow (A,Z-2) + 2e^{+} +2\nu
\end{equation}

\begin{equation}
e^- + (A,Z) \rightarrow (A,Z-2) + e^{+} + \nu + X
\end{equation}

\begin{equation}
e^- + e^- + (A,Z)  \rightarrow (A,Z-2)+ 2\nu + 2X
\end{equation}

These processes are not forbidden by any conservation laws, and their observation is interesting 
from the point of view of investigating nuclear-physics aspects of double-beta decay. Processes 
(6) and (7) are quite strongly suppressed because of low phase-space volume, and investigating 
process (8) is very difficult because one only has low energy X-rays to detect.  In the case of 
double-electron capture, it is again interesting to search for transitions to the excited states 
of daughter nuclei \cite{BAR94}, which are easier to detect from an experimental point of view. 

A review of experimental investigations of $2\beta^+$, $\beta^+$EC and ECEC processes is presented 
in \cite{BAR04}.  It has to be mentioned here that a ``positive``
result was obtained in a geochemical 
experiment with $^{130}$Ba where the ECEC$(2\nu)$ process was detected with a half-life of $(2.2 
\pm 0.5)\cdot 10^{21}$ y \cite{MES01}.

The present work is dedicated to search for different modes of $\beta^+$EC and ECEC processes in 
$^{74}$Se.  

\section{Experimental}

The experiment has been performed in the Modane Underground Laboratory (depth of 4800 m w.e.). The 
natural selenium powder sample was measured using a 400 cm$^3$ low-background HPGe detector.

The Ge spectrometer was composed of p-type crystals. For the HPGe detector the cryostat, the 
endcap and the main mechanical parts were made of very pure Al-Si alloy. The cryostat had a J-type 
geometry to shield the crystal from radioactive impurities in the dewar. The passive shielding 
consisted of 4 cm of Roman-era lead and 3-10 cm of OFHC copper inside 15 cm of ordinary lead. To 
remove $^{222}$Rn gas, one of the main sources of the background, a special effort was made to 
minimize the free space near the detector. In addition, the passive shielding was enclosed in 
an aluminum box flushed with high-purity nitrogen.

The electronics consisted of currently available spectrometric amplifiers and a 8192 channel ADC. 
The energy calibration was adjusted to cover the energy range from 50 keV to 3.5 MeV, 
and the energy resolution was 2.0 keV for the 1332-keV line of $^{60}$Co. The 
electronics were stable during the experiment due to the constant conditions in the laboratory 
(temperature of $\approx 23^\circ$ C, hygrometric degree of $\approx 50$\%).  A daily check on the 
apparatus assured that the counting rate was statistically constant.

The sample of natural selenium powder was placed in a circular plastic box and put on the endcap 
of the HPGe detector. The mass of the powder was 563 g, 4.69 g of which was $^{74}$Se (as the 
natural abundance is 0.89\%). The duration of measurement was 436.56 hours.

A search for different $\beta^+$EC and ECEC processes in $^{74}$Se was carried out using the 
germanium detector to look for $\gamma$-ray lines corresponding to these processes. Hereinafter, 
$Q'$ is the effective $Q$-value defined as $Q'=\Delta {\rm M} - \epsilon_1-\epsilon_2$ where 
$\Delta M$ is the difference of parent and daughter atomic masses, $\epsilon_i$ is an electron 
binding energy in a daughter nuclide.

The ECEC$(0\nu)$ transitions were considered for three cases of electron captures as is shown 
in Fig.~1.\\ First, two electrons are captured from the L shell. In this case, $Q'$ is equal to 
$\sim 1206.9$ keV and three transitions are investigated, i.e. \\
1) to the second $2^+$ level of $^{74}$Ge (1204.2 keV), accompanied by 595.8 keV and 608.4 keV 
de-excitation $\gamma$-quanta (66\%), or a 1204.2 keV de-excitation $\gamma$-quantum (34\%); \\
2) to the first $2^+$ level of $^{74}$Ge (595.8 keV), accompanied by a 611.1 keV bremsstrahlung 
$\gamma$-quantum and a 595.8 keV de-excitation $\gamma$-quantum; \\
3) to the ground state of $^{74}$Ge, accompanied by a 1206.9 keV bremsstrahlung $\gamma$-quantum. 
\\
Second, one electron is captured from the K shell, another from the L shell.  In this case, $Q'$ 
is equal to $\sim 1197.2$ keV and two transitions are investigated, i.e. \\
1) to the first $2^+$ level of $^{74}$Ge, accompanied by a 601.4 keV bremsstrahlung 
$\gamma$-quantum and a 595.8 keV de-excitation $\gamma$-quantum; \\
2) to the ground state of $^{74}$Ge, accompanied by a 1197.2 keV bremsstrahlung $\gamma$-quantum. 
\\
Third, two electrons are captured from the K shell.  In this case, $Q'$ is equal to $\sim 1185.9$ 
keV and two transitions are investigated, i.e. \\
1) to the first $2^+$ level of $^{74}$Ge, accompanied by a 590.1 keV bremsstrahlung 
$\gamma$-quantum and a 595.8 keV de-excitation $\gamma$-quantum; \\
2) to the ground state of $^{74}$Ge, accompanied by a 1185.9 keV bremsstrahlung $\gamma$-quantum.

The $\beta^+$EC$(0\nu + 2\nu)$ transition is possible only to the ground state of $^{74}$Ge, 
accompanied by one positron which gives two annihilation $\gamma$-quanta.  The modes of $(0\nu$ 
and $2\nu)$ aren't distinguished with our technique.

The usual ECEC$(2\nu)$ transitions, accompanied by detectable $\gamma$-rays, are  \\
1) the transition to
 the second $2^+$ level of $^{74}$Ge with the $\gamma$-rays, 
595.8 keV (66\%), 608.4 keV (66\%) and 1204.2keV (34\%), \\
2) the transition to the first $2^+$ level of $^{74}$Ge with 
  one de-excitation $\gamma$-quanta, 595.8 keV. \\

\begin{figure*}
\begin{center}
\resizebox{0.75\textwidth}{!}{\includegraphics{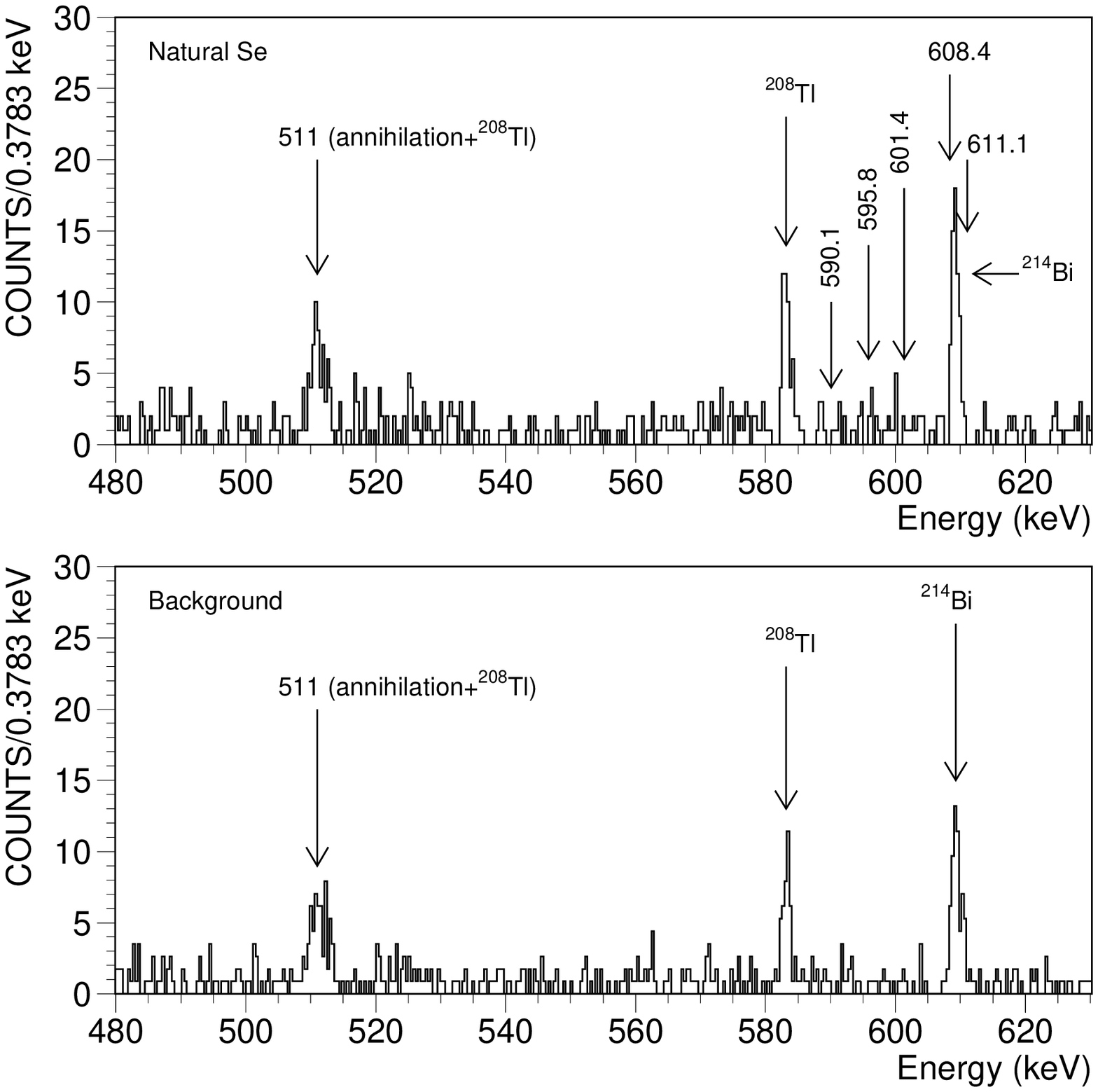}}
\resizebox{0.75\textwidth}{!}{\includegraphics{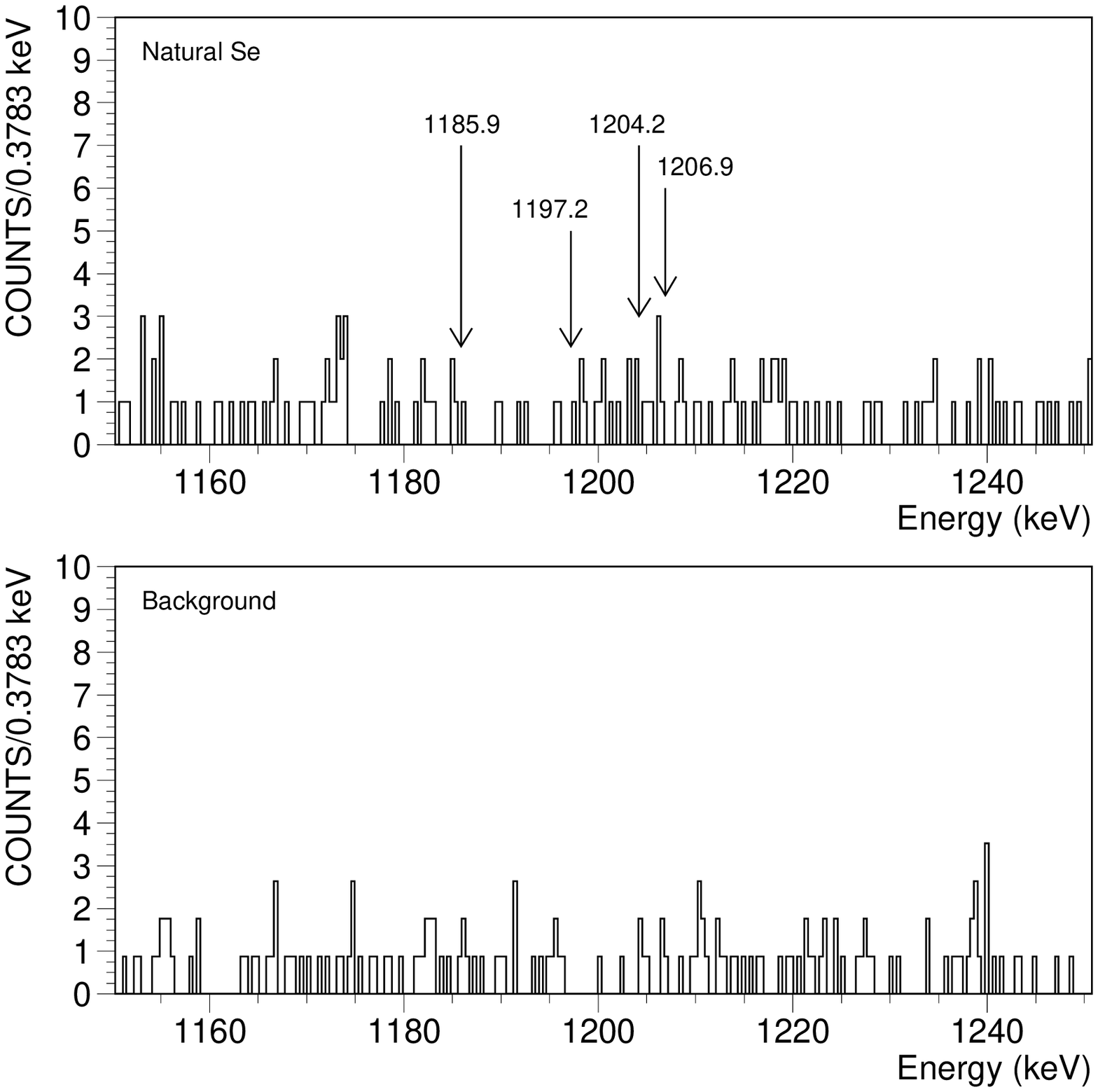}}
\caption{Energy spectra with natural Se and without Se (background) in the ranges of investigated $\gamma$-rays.}  
\label{fig_2}
\end{center}
\end{figure*}

The partial gamma-ray spectra in the energy ranges corresponding to the different decay modes of 
$^{74}$Se are shown in Fig.~2 together with the background (without Se powder) spectrum reduced to 
the measuring time of the experiment. In the Fig.~2 all visible peaks in experimental and 
background spectra are within one sigma error, i.e. they are background peaks. No extra events 
that are statistically significant, i.e. more than $3\sigma$ over background, are observed for the 
investigated energies.

The Bayesian approach \cite{PDG04} was used to estimate limits on transitions of $^{74}$Se to the 
ground and excited states of $^{74}$Ge. To construct likelihood functions every bin of the 
spectrum is supposed to have a Poisson distribution with its mean $\mu_i$ and the number of events 
equal to the content of this $i$th bin. The mean $\mu_i$ can be written in general form as

\begin{equation}
\mu_i = N\sum_{m} {\varepsilon_m a_{mi}} + \sum_{k}
{P_k a_{ki}} + b_i
\end{equation}

The first term in (9) describes the contribution of the investigated process that may have a few 
$\gamma$-lines contributing appreciably to the $i$th bin; the parameter $N$ is the number of 
decays, $\varepsilon_m$ is the detection efficiency of the $m$th $\gamma$-line of this transition 
and $a_{mi}$ is the contribution of $m$th line to the $i$th bin. For low-background measurements a 
$\gamma$-line may be taken to have a gaussian shape. The second term gives contributions of 
background $\gamma$-lines; here $P_k$ is the area of the $k$th $\gamma$-line and $a_{ki}$ is its 
contribution to the $i$th bin. The third term represent the so-called ``continuous background'' 
$b_i$ which has been selected as a straight-line fit after rejecting all peaks in the region of 
interest.  We have selected this region as the peak investigated $\pm$ 30 standard deviations 
($\approx$ 20 keV). The likelihood function is the product of probabilities for selected bins.  
Normalizing this to 1 over parameter $N$, it becomes the probability density function for $N$, 
which is used to calculate limits for $N$.  To take into account errors of $\gamma$-line shape 
parameters, peak areas, and other factors, one should multiply the likelihood function by the 
error probability distributions for those values, and integrate on those, providing the averaging 
probability density function for $N$.

In our case, we have taken into consideration only errors of peak areas for the 511-keV 
annihilation peak and for the 609.3-keV gamma-line from $^{214}$Bi, supposing other errors are 
negligible.  Limits have been calculated for different combinations of $\gamma$-lines 
corresponding to the transitions under study.  The best results are given in Table 1. 

The photon detection efficiency for each investigated process has been computed with the CERN 
Monte Carlo code GEANT3.21. 
Special calibration measurements with radioactive sources and powders containing well-known 
$^{226}$Ra activities confirmed that the accuracy of these efficiencies is about 10\%.

\begin{table}[ht]
\label{Table1}
\caption{The limits on double-beta processes in $^{74}$Se. The second column presents $\gamma$-ray 
energies in keV and their efficiencies used to estimate half-lives. Limits on half-lives $T_{1/2}$ 
are given at a confidence level of 90\%. $^{*)}$ For transition with irradiation of $e^+e^-$ pair - see footnote $^{3)}$.
$^{**)}$ This transition is forbidden and the limit is presented here just for comparison.}
\begin{center}
\begin{tabular}{lll}
\hline
Transitions &  $\gamma$-ray(efficiency) & $T_{1/2}, 10^{19}$ y \\
\hline
${\rm L}^1{\rm L}^2, Q'=1206.9$ keV,  &  595.8 keV (1.88\%) \\
ECEC$(0\nu)$ to the $2^+_2$(1204.2-keV) level of $^{74}$Ge                             &  608.4 keV (1.84\%) & 0.55 \\
                                                                                                       &  1204.2 keV (0.757\%) \\
\hline
${\rm L}^1{\rm L}^2, Q'=1206.9$ keV,   & 595.8 keV (2.76\%) \\
ECEC$(0\nu)$ to the $2^+_1$(595.8-keV) level of $^{74}$Ge                                & 611.1 keV (2.70\%) & 1.30 \\
\hline
${\rm L}^1{\rm L}^2, Q'=1206.9$ keV, \\
ECEC$(0\nu)$ to the ground state of $^{74}$Ge &  1206.9 keV (2.13\%) & 0.41 \\
\hline
${\rm K}^1{\rm L}^2, Q'=1197.2$ keV,  &  595.8 keV (2.75\%) \\
ECEC$(0\nu)$ to the $2^+_1$(595.8-keV) level of $^{74}$Ge                              &  601.4 keV (2.71\%) & 1.12 \\
\hline
${\rm K}^1{\rm L}^2, Q'=1197.2$ keV, \\
ECEC$(0\nu)$ to the ground state of $^{74}$Ge &  1197.2 keV (2.13\%) & 0.64 \\
\hline
${\rm K}^1{\rm K}^2, Q'=1185.9$ keV,  &  595.8 keV (2.75\%) \\
ECEC$(0\nu)$ to the $2^+_1$(595.8-keV) level of $^{74}$Ge                   &   590.1 keV (2.76\%) & 1.57 \\
\hline
${\rm K}^1{\rm K}^2, Q'=1185.9$ keV, & 511 keV (6.74\%) & 0.19$^{*)}$ \\
ECEC$(0\nu)$ to the ground state of $^{74}$Ge                       &  1185.9 keV (2.15\%) & 0.62$^{**)}$ \\
\hline
$\beta^+$EC$(0\nu + 2\nu)$ transition \\
to the ground state of $^{74}$Ge                                            &  511 keV (6.74\%) & 0.19 \\ 
\hline
ECEC$(2\nu)$ to the ($2^+_2)$(1204.2-keV) level of $^{74}$Ge &  595.8 keV (1.88\%) \\
                                                                           &  608.4 keV (1.84\%) & 0.55 \\
                                                                           &  1204.2keV (0.757\%) \\
\hline
ECEC$(2\nu)$ to the $2^+_1$(595.8-keV) level of $^{74}$Ge &  595.8 keV (3.13\%) & 0.77 \\
\hline
\end{tabular}
\end{center}
\end{table}

\section{Discussion} 
The obtained results are presented in Table 1. It is necessary to stress that 
$^{74}$Se has never been investigated before and all results presented here are obtained for 
the first time.  Neither has 
this isotope been investigated theoretically; thus, there are no predictions with which to 
compare.  Nevertheless, we will try to estimate the significance of the obtained results and 
the possibility to increase the sensitivity of this type of experiments in the future.

It is clear that $^{74}$Se is not a very good candidate to search for $\beta^+$EC$(2\nu)$ and 
ECEC$(2\nu)$ processes and chance of detecting these decays is very small (even taking into 
account a possible increase in sensitivity: see below). Nevertheless, the obtained experimental 
limits are quite interesting as they prohibit some unexpected (exotic) processes.

Concerning the ECEC$(0\nu)$ processes, the main hope is to observe a resonant transition to the 
1204.2-keV excited state of $^{74}$Ge. In fact, the experimental spectrum has some excess
of events in the ranges of 595.9 keV (1.8 $\sigma$). Most probably that this excess of events is 
connected with cosmogenic isotope $^{74}$As (T$_{1/2}$ = 17.8 d).
This isotopes is produced on the surface in reaction $^{74}$Se(n,p)$^{74}$As. Fortunately half-life
of this isotope is quite short and it can not be a serious background for future 
underground experiments (one just has to wait a few months before starting the measurement).

It would therefore be better to investigate a larger quantity of $^{74}$Se for a longer time. 
Further, it is necessary to precisely determine the $\Delta M$ value for $^{74}$Se and $^{74}$Ge; 
this is a realistic task (in Ref. \cite{FRE05}, the possibility of measurements with an accuracy 
$\sim 200$ eV is discussed).  A theoretical investigation of this transition and an estimate of 
the rate of ECEC$(0\nu)$ process for different channels are needed.  Experimental possibilities 
are as follows: \\ 
1) With 1 kg of enriched $^{74}$Se in the setup described in the preceding 
section, the sensitivity after one year of measurement will be $\sim 3\cdot 10^{21}$ y. \\
2) To use $\gamma-\gamma$ coincidence technique that uses two HPGe detectors to observe cascade of 
the two $\gamma$ rays 
(595.8 keV and 608.4 keV). Using detector described in 
\cite{DEB05} with 1 kg of enriched $^{74}$Se and after one year of measurement sensitivity 
will be on the level $\sim 3\cdot 10^{21}$ y. \\
3) With 
200 kg of enriched $^{74}$Se using an installation such as GERDA \cite{ABT04} or MAJORANA 
\cite{MAJ03,AAL05} (where 500 kg of low-background HPGe detectors are planned to be used), the 
sensitivity after 10 years of measurement may reach $\sim 10^{26}$ y. \\

4) To use TGV type detector \cite{STE06} which can reach sensitivity $\sim 10^{20}$ y with
10 g of $^{74}$Se for one year of measurement. \\
It is emphasized that these 
types of measurements can be done with some other isotopes as well.  In Table 2, the most 
interesting of such isotopes are presented along with some relevant properties. One of the most 
promising candidates is $^{112}$Sn.

\begin{table}[ht]
\label{Table3}
\caption{The prospective isotopes. Here A is isotopic abundance, $\Delta M$ is the atomic mass 
difference in keV of parent and daughter nuclei, E* is the energy in keV of a candidate excited 
state of the daughter nuclide (with its spin and parity in parenthesis) for a resonant transition, 
and $E_K, E_{L1}, E_{L2}, E_{L3}$ are the energies of K, L1, L2, and L3 shells of a daughter 
nuclide in keV.}
\begin{center}
\begin{tabular}{ccccccccc}
\hline
Nuclei &	A, \% & $\Delta M$ & E* & $E_K$ & $E_{L1}$ &  $E_{L2}$ & $E_{L3}$  \\
\hline
$^{74}$Se  & 0.89 & $1209.7\pm 2.3$ & $1204.2 (2^+)$ & 11.10 & 1.41 & 1.23 & 1.22 \\ 
\hline
$^{78}$Kr    & 0.35 & $2846.4\pm 2.0$ & $2838.9 (2^+)$ & 12.66  & 1.65 & 1.47 & 1.43 \\
\hline
$^{96}$Ru   & 5.52 & $2718.5\pm 8.2$ & $2700 (?)  $       & 20     & 2.86 & 2.62 & 2.52 \\
\hline
$^{106}$Cd & 1.25 & $2770\pm 7.2$   & $2741.0 (1,2^+)$  & 24.35 & 3.60 & 3.33 & 3.17 \\
\hline
$^{112}$Sn & 0.97  & $1919.5\pm 4.8$  & $1871.0 (0^+)$  & 26.71 & 4.02 & 3.73 & 3.54 \\
\hline
$^{130}$Ba & 0.11 & $2617.1\pm 2.0$  & $2608.42 (?) $& 34.56 & 5.44 & 5.10 & 4.78 \\
\hline
$^{136}$Ce & 0.20 & $2418.9\pm 13$  & $2399.87 (1^+,2^+)$ & 37.44  & 5.98  & 5.62 & 5.25 \\
                 &         &                          &  $2392.1 (1^+,2^+)   $ \\ 
\hline
$^{162}$Er  & 0.14 & $1843.8\pm 5.6$ & $1745.71 (1^+)$ & 53.79  & 9.05  & 8.58 & 7.79 \\
\hline
\end{tabular}
\end{center}
\end{table}

\section{Conclusion}

For the first time, limits on double-beta processes in $^{74}$Se have been obtained. Some other 
prospective isotopes where the resonance mechanism can be realized are presented. It is 
demonstrated both that, in future larger-scale experiments, the sensitivity to ECEC($0\nu$) 
processes for such isotopes can be on the order of $10^{26}$ y and that, under resonant 
conditions, this decay will be competitive with $0\nu\beta\beta$ decay.

\section*{Acknowledgement}
The authors would like to thank the Modane Underground Laboratory staff for their technical 
assistance in running the experiment. We are very thankful to Dr. V. Tretyak 
for his useful remarks. Portions of this work were supported by a grant from INTAS 
(no 03051-3431).

 --------------------------------------------------------------

\end{document}